\documentclass[letter,twocolumn]{jpsj3}
\usepackage{amssymb}
\usepackage{txfonts}
\usepackage{array}
\usepackage{chngpage}
\usepackage[dvipdfmx]{graphicx}
\usepackage{dcolumn}
\usepackage{bm}

\newcolumntype{C}[1]{>{\hfil}m{#1}<{\hfil}}

\title{Inverse Design of Strongly Localized Topological $\pi$ Modes in One-Dimensional Nonperiodic Systems}

\author{Fumitatsu Iwase}
\inst{Department of Physics, Tokyo Medical University\\
 6-1-1 Shinjuku, Shinjuku-ku, Tokyo 160-8402, Japan}

\date{\today}

\abst{
This study investigates the spatial confinement of topological $\pi$-modes in one-dimensional chiral-symmetric systems.
In conventional periodic and quasiperiodic structures, edge-mode wave functions inevitably penetrate the bulk.
To suppress this, inverse design of a potential sequence is performed using a generative model under a global topological constraint.
The generated sequence reveals a characteristic structure consisting of a topological boundary layer and a macroscopic S-dense domain, leading to enhanced confinement ($\xi=0.85$) while preserving topology.
Based on the physical principle extracted from this result, a minimal heterostructure composed of only two S-blocks is manually constructed, which further reduces the localization length to $\xi=0.75$.
These results provide a compact design principle for strongly localized topological states.
}

\begin{document}
\maketitle

Topological insulators have provided robust edge and surface states protected by global topological invariants~\cite{HasanKane2010, Qi2011, Kitagawa2010}.
These states have been extensively studied in condensed matter systems, particularly in periodic lattices.
In such systems, the spatial confinement of edge states is largely determined by the bulk energy gap.
The edge-mode wave function decays exponentially into the bulk, with a decay length set by the gap size, resulting in a finite penetration of the edge state~\cite{Asboth2016}.

Recently, topological phenomena have been explored beyond periodic systems, particularly in aperiodic structures such as quasiperiodic Fibonacci lattices~\cite{Kohmoto1983}.
Despite the absence of translational symmetry, these systems exhibit characteristic features including fractal energy spectra and multiple gaps, which can enhance wave localization through interference effects~\cite{Kraus2012}.

However, while topological edge states are known to be robust against random disorder~\cite{Cedzich2018, Hetenyi2021}, quasiperiodic systems based on deterministic substitution rules still exhibit a finite penetration of the edge-mode wave function into the bulk~\cite{Kraus2012,Verbin2013}.
In particular, complete suppression of the evanescent tail remains difficult to achieve in such structures.

\begin{figure}[b]
\begin{center}
\includegraphics[width=8.5cm]{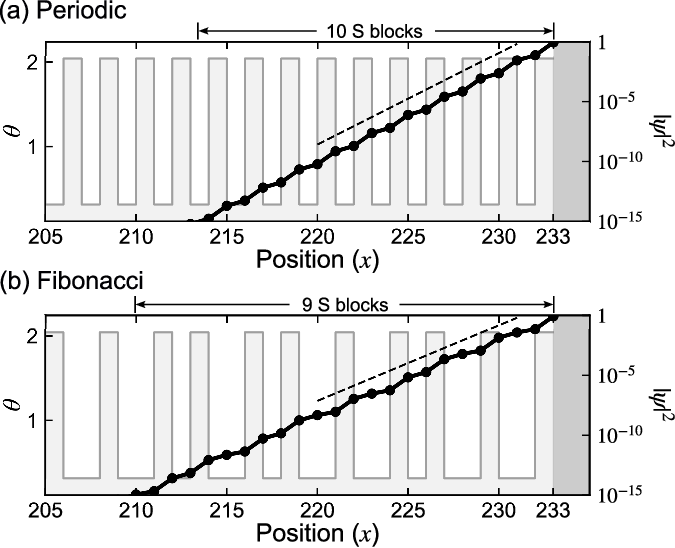}%
\caption{\label{fig:periodic_fibonacci}
Spatial distributions of the coin parameters and decay curves of the probability density $|\psi|^2$ for the $\pi$-mode for (a) the periodic lattice and (b) the Fibonacci lattice, with enlarged views of the rightmost 28 sites.
The quantity $|\psi|^2$ is plotted on a logarithmic scale, and the dashed lines represent the slopes of the fitting curves indicating exponential decay.
The binary coin parameters correspond to the rotation angles $\theta_\mathrm{W}=0.1\pi$ and $\theta_\mathrm{S}=0.65\pi$.
The brackets shown at the top of each panel indicate the spatial range and the corresponding number of S blocks required for the probability density to reach $10^{-15}$.
}
\end{center}
\end{figure}

Figure~\ref{fig:periodic_fibonacci} shows the spatial distributions of the binary coin parameters for the periodic and Fibonacci sequences employed in the present quantum-walk model, together with the corresponding localization profiles of the $\pi$-mode.
The $\pi$-mode localized at the right edge exhibits an exponential decay, characterized by the localization length $\xi$, defined through $|\psi|^2 \propto \exp(2x/\xi)$.
The corresponding values are $\xi=1.14$ for the periodic sequence and $\xi=1.38$ for the Fibonacci sequence, indicating that the evanescent tail is not sufficiently suppressed in either case.
A more detailed discussion of these localization properties is given later.

To address this limitation, machine-learning approaches have recently been applied to inverse design problems of potential sequences~\cite{Ma2021,Molesky2018,Pilozzi2018,Ghafiani2024,Jiang2025}.
In particular, generative models such as generative adversarial networks (GANs) provide a framework for exploring complex configuration spaces and identifying candidate structures~\cite{Ma2021,Liu2018}.
By learning from a dataset of stochastically generated samples, such models can capture underlying structural features and generate configurations beyond those accessible by simple random search or local optimization.
This approach enables a systematic search for structures that satisfy both physical constraints and desired functional properties.

\begin{figure*}[t]
\begin{center}
\includegraphics[width=17.5cm]{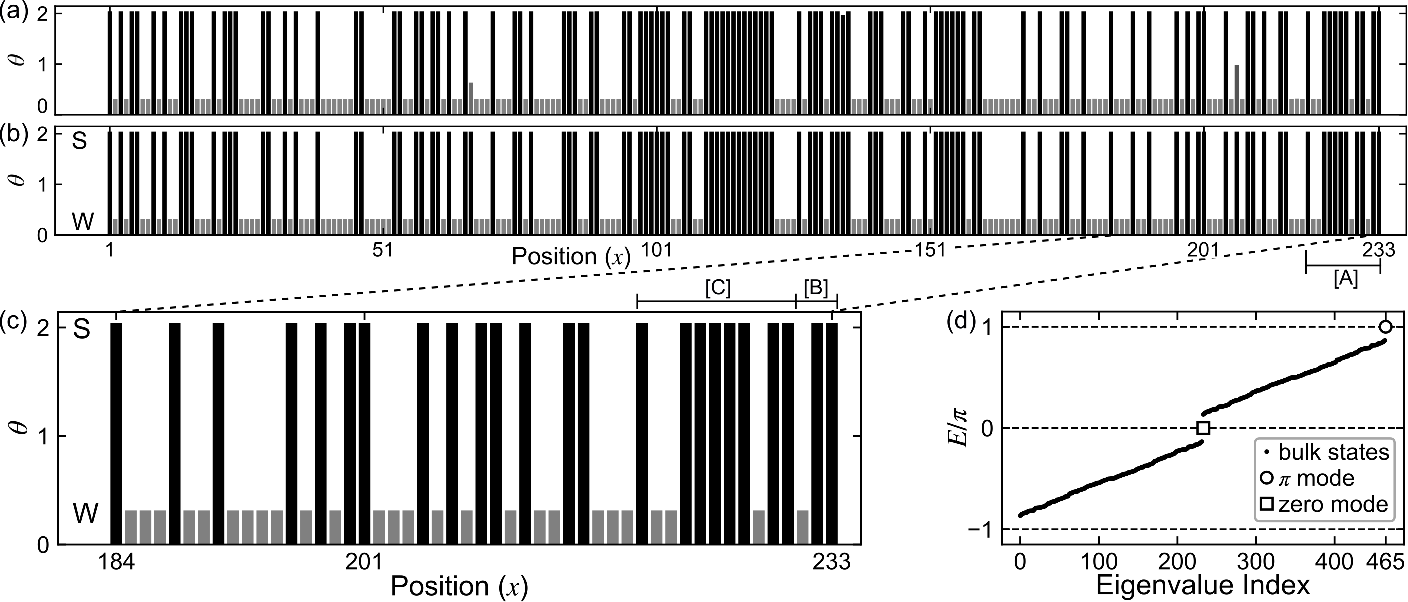}%
\caption{\label{fig:generated_sequence}
Spatial distributions of the generated coin parameters and the corresponding quasienergy spectrum.
(a) Raw output sequence from the generative model.
(b) Sequence binarized from (a) with a 0.5 threshold.
Region [A] denotes the hierarchical edge heterostructure.
(c) Enlarged view of the rightmost 50 sites of (b).
Regions [B] and [C] indicate the topological host layer and the internal S-dense domain, respectively.
(d) Quasienergy spectrum.
The zero and $\pi$ modes, highlighted by open square ($\square$) and open circle ($\bigcirc$), are clearly isolated from the dense bulk continuum.
The total number of eigenvalues is $2N$, due to the two internal coin states.
}
\end{center}
\end{figure*}

In this Letter, inverse design of a one-dimensional chiral-symmetric topological system is performed using a generative model, with the aim of enhancing the localization of the topological $\pi$-mode.
The generated sequence exhibits a nonperiodic structure characterized by a hierarchical edge arrangement, including a local domain that effectively separates the edge region from the bulk ($\xi=0.85$).
Analysis of the structural features revealed by the machine-learning-generated sequence further leads to the manual construction of a minimal heterostructure.
This sequence achieves even stronger localization than all other configurations examined in the present study.


A discrete-time quantum walk is considered on a one-dimensional lattice~\cite{Aharanov1993,Kitagawa2010,Asboth2012} of size $N=233$ under open (reflective) boundary conditions.
The Hilbert space is given by the tensor product of the position basis $\{|x\rangle\}$ ($x=0,\dots,N-1$) and the internal coin states $\{|L\rangle, |R\rangle\}$, representing left- and right-moving components, respectively.
The time-evolution operator is defined as $U=SC$, where the shift operator $S$ and the spatially modulated coin operator $C$ are given by:
\begin{eqnarray}
    S &=& \sum_x \left( |x-1\rangle \langle x| \otimes |L\rangle \langle L| 
    + |x+1\rangle \langle x| \otimes |R\rangle \langle R|\right),\\
    C &=& \sum_x |x\rangle \langle x|\otimes \begin{pmatrix}
        \cos\theta_x & \sin\theta_x \\
        -\sin\theta_x & \cos\theta_x
    \end{pmatrix}.
\end{eqnarray}
The rotation angle $\theta_x$ takes either a weak ($\theta_\mathrm{W}=0.1\pi\approx 0.314$, denoted by W) or strong ($\theta_\mathrm{S}=0.65\pi\approx 2.04$, denoted by S) potential.
As reference systems, a periodic sequence and Fibonacci sequence (13th generation, generated via A $\rightarrow$ AB, B $\rightarrow$ A) are considered.
For the present choice of parameters, both the periodic and Fibonacci sequences are confirmed to be topologically nontrivial from the corresponding real-space topological invariant~\cite{Cedzich2022}.
For the periodic sequence, this nontrivial character is also consistent with the Zak phase~\cite{Asboth2012}.
In the following, we focus on the topological $\pi$-mode localized at the right edge of the system.


To maximize the localization of the right-edge topological $\pi$-mode $|\psi\rangle$ (obtained by diagonalizing $U$), the first moment is defined as 
\begin{equation}
\mu_\mathrm{R} = \sum_{x=0}^{N-1} \left( (N-1) -x \right) |\psi(x)|^2,
\end{equation}
and the corresponding normalized fitness score is given by 
\begin{equation}
F = \exp (-\mu_\mathrm{R} / \alpha),
\end{equation}
with $\alpha = 2.0$.
A score of $F=1.0$ indicates perfect confinement ($\mu_\mathrm{R} = 0$).

For automated inverse design of a nonperiodic heterostructure, a generative adversarial network (GAN) is employed~\cite{Ma2021,Liu2018}.
An initial set of 10,000 random sequences was generated for the training dataset.
The fitness score $F$ was then evaluated for each sequence, and the top 10\% (1,000 elite sequences) were selected as the training data for the GAN.

To prevent the generation of topologically trivial phases, the topological invariant (winding number $W$) of the generated sequences was evaluated using real-space analysis~\cite{Cedzich2022}.
By filtering the training data and implicitly guiding the generator based on this invariant, the machine-learning model learns the global topological constraints.

While simpler heuristic algorithms may also yield localized solutions, the GAN is employed here as an efficient tool for exploring the vast discrete configuration space ($2^N$).
Based on the characteristic features identified from the generated sequences, a minimal configuration was subsequently constructed to further enhance the localization.


To investigate the optimization strategy of the generative model, the overall potential profile of the generated lattice, shown in Fig.~\ref{fig:generated_sequence}(a), is examined. 
This sequence is obtained by feeding noise into the trained generator, yielding continuous values between 0 and 1, which are then linearly mapped onto the coin parameter range.
Although no explicit binary constraint was imposed during training, the resulting sequence exhibits an almost binary profile.
Figure~\ref{fig:generated_sequence}(b) shows the binarized sequence obtained by applying a threshold of 0.5 to the continuous data.
In the following, the binarized sequence is analyzed for clarity and ease of comparison with other structures.
It has been confirmed that this procedure does not alter the essential features of the generated configuration.
The real-space topological invariants evaluated from both ends remain nontrivial.


A characteristic feature of the generated sequence is found in the microscopic arrangement near the right boundary.
Figure~\ref{fig:generated_sequence}(c) shows an enlarged view of the rightmost 50 sites of Fig.~\ref{fig:generated_sequence}(b).
Near the edge, the sequence terminates as $\cdots$SWWSSSSSWSSWSS, as indicated by region [A] in Fig.~\ref{fig:generated_sequence}(b). 
This edge portion forms a functional heterostructure, which we refer to as a hierarchical edge heterostructure, consisting of a topological host layer (WSS, region [B]) that accommodates the topological $\pi$-mode, followed by an S-dense domain (SWWSSSSSWSS, region [C]).
Beyond this S-dense domain, the residual probability density decays gradually due to the small band-gap of the W potentials.
While a higher density of S potentials near the edge can enhance confinement, the generated structures show a distinct spatial organization, where a topological host region and a localized domain are systematically arranged.

The quasienergy spectrum for the generated sequence is shown in Fig.~\ref{fig:generated_sequence}(d), where the quasienergy $E$ is derived from the eigenvalue $\lambda$ ($\lambda = e^{-iE}$) of the time-evolution operator $U$.
Owing to the aperiodic structure, the bulk states are densely distributed over the quasienergy range, forming an almost continuous and nearly linear spectrum.
Despite this dense distribution, a local gap remains open around the edge-state energy, within which the topological $\pi$-mode is clearly isolated from the bulk states.

Similar microscopic edge structures are reproducibly obtained in multiple independent runs, with only minor variations in the detailed sequence, indicating that the characteristic motif is robust against the stochastic nature of the generation process.
This type of structure is not observed in conventional periodic or quasiperiodic sequences.

\begin{figure}
\begin{center}
\includegraphics[width=8.5cm]{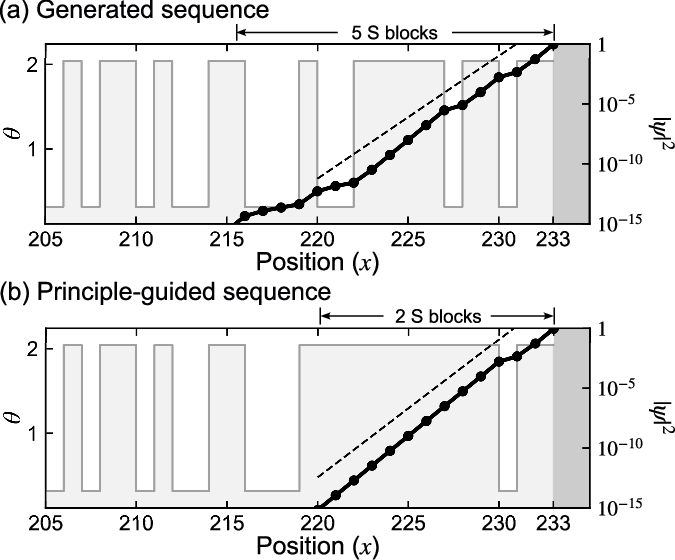}%
\caption{\label{fig:localization_length}
Decay curves of the probability density $|\psi|^2$ for the $\pi$-mode localized at the right boundary.
The panels show the results for the (a) generated sequence, and (b) principle-guided sequence constructed manually based on the generated design.
The format of the figure is identical to that in Fig.~\ref{fig:periodic_fibonacci}.
}
\end{center}
\end{figure}

Edge confinement is quantitatively evaluated by comparing the probability density $|\psi|^2$ with those of the standard reference models [Figs.~\ref{fig:periodic_fibonacci} and ~\ref{fig:localization_length}(a)].
The periodic alternating lattice [Fig.~\ref{fig:periodic_fibonacci}(a)] exhibits an exponential decay consistent with the a uniform bulk gap.
By fitting this decay with $|\psi |^2 \propto \exp(2x/\xi)$, the localization length is obtained as $\xi = 1.14$.
The quasiperiodic Fibonacci lattice [Fig.~\ref{fig:periodic_fibonacci}(b)] shows a slower overall decay compared with the periodic case, reflecting its multi-gap structure, but also exhibits oscillatory tails, resulting in a localization length $\xi = 1.38$.

In contrast, the machine-learning-generated lattice [Fig.~\ref{fig:localization_length}(a)] exhibits a decay profile similar to that of the  host region near the edge, followed by a sharp drop at the position of the SSSSS domain.
The local decay length estimated in this region is $0.85$, which is shorter than those of the periodic and Fibonacci cases.
This corresponds to a reduction of approximately 25\% and 38\%, respectively.
%
This result suggests that, while preserving the global topological phase, introducing an appropriately designed local domain can effectively suppress the penetration of the edge state into the bulk.

Next, the resource efficiency of these structures is compared. 
To suppress the probability density down to the numerical floor of $10^{-15}$, the periodic system requires approximately 10 consecutive S blocks, while the Fibonacci lattice requires 9. 
In contrast, the generated sequence achieves a comparable level of attenuation within 5 S blocks from the edge.
%
%
These findings suggest that inverse-designed structures can provide a compact route to realizing strongly localized topological edge states, offering a foundational blueprint for future robust topological devices.
Such highly resource-efficient nonperiodic sequences are difficult to identify solely by intuitive design, but can be further refined once the underlying structural principle is revealed.


Based on the structural principle extracted from the machine-learning-generated sequence, a simplified heterostructure was manually constructed.
Since placing three or more consecutive S potentials at the absolute right edge leads to a trivial phase ($W=0$), the topological boundary layer is fixed to an SS pair.
This principle-guided sequence consists of two distinct S blocks separated by a single W potential: an internal macroscopic S domain and an edge SS pair [Fig.~\ref{fig:localization_length}(b)].
This modification near the right edge leaves the quasienergy spectrum and the isolated $\pi$-mode essentially unchanged.
This minimal heterostructure exhibits even stronger attenuation, with a markedly sharper decay of the evanescent tail, and further reduces the localization length to $\xi=0.75$, exceeding both the conventional and the machine-learning-generated sequences.
This indicates that the enhanced confinement is determined not only by the presence of S-rich regions, but more critically by their specific spatial arrangement under the topological constraint.
These results suggest that the generative model can reveal a useful structural principle for strongly localized topological states.

Various machine-learning approaches, including reinforcement learning and evolutionary algorithms, have been explored for inverse design problems~\cite{Molesky2018,Melnikov2018,Wiecha2021}.
Here, the generative model plays a central role in revealing a physically interpretable design principle for strongly localized topological states, rather than merely serving as an optimization engine.
The results suggest that a hierarchical heterostructure, which enhances local contrast at the domain boundary, provides an efficient solution for maximizing edge-state localization while preserving the global topological invariant.
Furthermore, combining such edge-engineered structures with functional bulk systems, such as quasiperiodic lattices, may provide a route toward hybrid topological systems with enhanced control over wave localization and transport.
This perspective further suggests the usefulness of inverse design for uncovering nontrivial structural principles and design strategies in topological systems.

In this study, the inverse design of a one-dimensional chiral-symmetric topological system was performed using a generative model, with the aim of maximizing the localization of the topological $\pi$-mode.
The generated sequence exhibits a nonperiodic structure characterized by a hierarchical edge arrangement.
Quantitative analysis shows that this structure achieves a shorter localization length than conventional periodic and Fibonacci systems, leading to strong suppression of the wave function with a reduced number of strong potential blocks.
These results suggest that, while the global topological invariant ensures the existence of edge states, their spatial localization can be controlled independently through local structural design.
Furthermore, by extracting the structural principle revealed by the generated sequence, a minimal sequence that achieves even stronger localization than the machine-learning-generated configuration was manually constructed.
The present approach provides a compact route for designing strongly localized topological states and highlights the usefulness of inverse design in uncovering nontrivial structural principles.


\begin{thebibliography}{99}%
\item[] $^*$E-mail: iwasef@tokyo-med.ac.jp
\bibitem{HasanKane2010} M. Z. Hasan and C. L. Kane, Rev. Mod. Phys. \textbf{82}, 3045 (2010).
\bibitem{Qi2011} X.-L. Qi and S.-C. Zhang, Rev. Mod. Phys. \textbf{83}, 1057 (2011).
\bibitem{Kitagawa2010} T. Kitagawa, M. S. Rudner, E. Berg, and E. Demler, Phys. Rev. A \textbf{82}, 033429 (2010).
\bibitem{Asboth2016} J. K. Asb\'oth, L. Oroszl\'any, and A. P\'alyi, \textit{A Short Course on Topological Insulators} (Springer, 2016).
\bibitem{Kohmoto1983} M. Kohmoto, L. P. Kadanoff, and C. Tang, Phys. Rev. Lett. \textbf{50}, 1870 (1983).
\bibitem{Kraus2012} Y. E. Kraus, Y. Lahini, Z. Ringel, M. Verbin, and O. Zilberberg, Phys. Rev. Lett. \textbf{109}, 106402 (2012).
\bibitem{Cedzich2018}
C. Cedzich, T. Geib, F. A. Gr\"{u}nbaum, C. Stahl, L. Vel\'{a}zquez, A. H. Werner, and R. F. Werner, Ann. Henri Poincar\'{e} \textbf{19}, 325 (2018).
\bibitem{Hetenyi2021} B. Het\'enyi, S. Parlak, and M. Yahyavi, Phys. Rev. B \textbf{104}, 214207 (2021).
\bibitem{Verbin2013} M. Verbin, O. Zilberberg, Y. E. Kraus, Y. Lahini, and Y. Silberberg, Phys. Rev. Lett. \textbf{110}, 076403 (2013).

\bibitem{Ma2021} W. Ma, Z. Liu, Z. A. Kudyshev, A. Boltasseva, W. Cai, and Y. Liu, Nat. Photonics \textbf{15}, 77 (2021).
\bibitem{Molesky2018} S. Molesky, Z. Lin, A. Y. Piggott, W. Jin, J. Vuckovi\'{c}, and A. W. Rodriguez, Nat. Photonics \textbf{12}, 659 (2018).
\bibitem{Pilozzi2018} L. Pilozzi, F. A. Farrelly, G. Marcecci, and C. Conti, Commun. Phys. \textbf{1}, 57 (2018).
\bibitem{Ghafiani2024} M. E. Ghafiani, M. Elaouni, S. Khattou, Y. Rezzouk, M. Amrani, O. Marbouh, M. Boutghatin, A. Talbi, E. H. E. Boudouti, and B. Djafari-Rouhani, Phys. Wave Phenom. \textbf{32}, 48 (2024).
\bibitem{Jiang2025} Z. Jiang, Y. Wang, B. Ji, G. He, and C. Jiang, ACS Photonics \textbf{12}, 2566 (2025).
\bibitem{Liu2018} Z. Liu, D. Zhu, S. P. Rodrigues, K. Lee, and W. Cai, Nano Lett. \textbf{18}, 6570 (2018).

\bibitem{Aharanov1993} Y. Aharonov, L. Davidovich, and N. Zagury, Phys. Rev. A \textbf{48}, 1687 (1993).
\bibitem{Asboth2012} J. K. Asb\'oth, Phys. Rev. B \textbf{86}, 195414 (2012).

\bibitem{Cedzich2022} C. Cedzich, T. Geib, F. A. Gr\"{u}nbaum, L. Vel\'{a}zquez, A. H. Werner, and R. F. Werner, Commun. Math. Phys. \textbf{389}, 31 (2022).
%
\bibitem{Melnikov2018} A. A. Melnikov, H. P. Nautrup, M. Krenn, and H. J. Briegel, Proc. Natl. Acad. Sci. USA \textbf{115}, 1221 (2018).
\bibitem{Wiecha2021} P. R. Wiecha, A. Arbouet, C. Girard, and O. L. Muskens, Photonics Res. \textbf{9}, BB182 (2021).


\end{thebibliography}

\end{document}